\UseRawInputEncoding
\documentclass[preprint,dvips,numbers,sort&compress]{elsarticle}
\usepackage{ifpdf}
\usepackage{hyperref}
\usepackage{amsmath}
\usepackage{amssymb}
\usepackage{amsfonts}
\usepackage{amsthm}
\usepackage{arydshln}
\usepackage[toc, page]{appendix}
\usepackage{vector}
\usepackage{graphicx}
\usepackage{subfigure}
\usepackage{color}
\usepackage{paralist}
\usepackage{multirow}
\usepackage{booktabs}
\usepackage{setspace}
\graphicspath{{zhou_eps/}}

\numberwithin{equation}{section}
\begin{document}
\title{ 
Wave-Particle Duality of Ultrasound: Acoustic Softening Explained by Particle Treatment of Ultrasonic Wave}
\author[a2]{Libin Yang}
\address[a2]{ 
School of Industrial Automation, Beijing Institute of Technology, Zhuhai, 519088, China
Email: 12261@bitzh.edu.cn}
\author[a1]{Lixiang Yang}
\address[a1]{Department of Mechanical and Aerospace Engineering, The Ohio State University, Columbus OH, 43210 Email: yang.1130@buckeyemail.osu.edu}
%
%

\begin{abstract} \noindent
When ultrasonic wave is irradiated on materials, a small static stress is required to get materials yielding and flowing. This is called acoustic softening effect, also known as Blaha effect for a long time. In the past, this effect was explained by several continuum scale or meso-scale solid mechanics theories such as stress superposition or energy superposition theory, or crystal/dislocation plasticity. Due to a lot of microscopic complexities happening inside the materials during ultrasonic vibration, fully understanding of acoustic softening effect is not easy. In this paper, traditional solid mechanics theory is expanded by introducing several concepts in semi-conductor physics.  Four new aspects were introduced to understand acoustic softening effect. Firstly, contrary to most existed work in acoustic softening research area which treats ultrasound as waves, it was treated as particles. Secondly, crystal/dislocation plastic theory was simplified to a single equation. Thirdly, concepts of photoelectric effect or photo-voltaic effect were introduced. Analogy of electron movement due to light wave and defect movement due to ultrasonic wave was illustrated. Particularly, as light wave is treated as photon, ultrasonic wave is treated as phonon in this paper. Fourthly, defects such as point defects or line defects are assumed to have certain bonding energy. Their bonding energies are assumed to be quantized or discontinuous.  The band gap theory used in photo-voltaic theory is embraced to understand defects movements in solid mechanics due to ultrasonic phonon. 

\end{abstract}
\maketitle
\section{Introduction}
\label{s:intro} \noindent
Ultrasonic vibration has found a lot applications in industries, such as ultrasonic cut, ultrasonic cleaning,  and ultrasonic welding \citep{cai_acousto-plastic_2006, siddiq_theoretical_2009, hu_impact_2017}. Ultrasonic wave is also widely used in biomedical areas such as ultrasonic imaging, ultrasonic therapy, and ultrasound in dentistry\citep{gallego-juarez_power_2023}. All of these applications are related to how to use and control the energy ultrasound carries. In spite of the widely known Acousto-Plastic Effect (APE) application in industries and biomedical applications, the basic understanding of the APE mechanism is far from complete. In ultrasonic cut and welding, it is known that ultrasound will make materials soft.   Experimental findings shows ultrasound vibration can dramatically make the crystalline materials soft without significant heating, which is referred as acoustic softening \citep{deng_study_2023}. This acoustic softening effect is temporary. It only appears during the application of the high-frequency vibration and will vanish as soon as the vibration stops.  It is shown that acoustic softening can more efficiently reduce yielding stress compared to heating softening. Acoustic energy is assumed to be absorbed only by lattice imperfection area (dislocations or grain boundaries) while heat is distributed homogeneously among all the atoms of the crystals.  Dislocation can be more easily moved by periodic stress than by constant stresses \citep{baker_dislocation_2004}. Therefore, APE is a more effective way to metal yielding, because a relatively small stress is needed and the effect is fast. Strain hardening process assisted by ultrasonic vibration is more effective as compared to that in conventional strain hardening processes. Experimental observations show the stress reduction by acoustic softening in different crystalline materials is proportional to the vibration amplitude \citep{izumi_effects_1966, sedaghat_ultrasonic_2019}. Since the amplitude of the imposed vibrations is proportional to the input energy, it is widely accepted that the amplitude of the vibrations is an important factor for the APE processes \citep{wang_acoustic_2016, cheng_acoustic_2023}. The frequency effect under ultrasonic vibration is in contradiction. But most researchers thought that frequency effect is negligible \citep{nevill_effect_1957, kirchner_plastic_1985}.  Moreover, most of research works acknowledged that the influence of
temperature on the APE is negligible.

In more than a half century, various theories have been proposed to describe the basic mechanisms of the APE process. From historic experimental development and observation, APE can depend on parameters like the frequency and amplitude of imposed vibration, and the temperature of environment. Several hypotheses have been proposed to explain experimental results. These theories can be classified into three groups: (i) the dislocation / potential-well hypothesis, (ii) the stress-superposition hypothesis, and (iii) the energy superposition hypothesis. But a comprehensive mathematical model is unavailable.  In the process of mathematical derivation, many empirical equations and assumptions were introduced but cannot easily be validated. A complete review of acoustic softening was given by Liu and Graff \citep{gallego-juarez_power_2023}. In chapter 13 of their book, more than 170 articles are collected spanning from early 1950s to 2023. Following the history line, some representative work in acoustic softening research area will be given in the following section.

Nevill \citep{nevill_effect_1957} reported that the decrease of yielding stress in tension in a low-carbon steel specimen is proportional to the vibrational amplitude and independent of the frequency of vibration in the frequency range of 15 kilo cycle to 80 kilo cycle. Study by Blaha and Langenecker \citep{blaha_plastizitatsuntersuchungen_1959} showed that sound waves would change the energy level of the absorbing dislocations. When a static stress is applied to the metal sample, the dislocation velocity will increase and plastic deformation would occur. Consequently, the apparent static stress which is needed for plastic deformation will be reduced. 

To further understand ultrasound induced dislocation movements, Langenecker \citep{langenecker_effects_1966} conducted a series of experiments on zinc, aluminum, beryllium, tungsten, low-carbon steel and stainless steel. He concluded that sound energy was absorbed only by lattice imperfections. But thermal energy was absorbed homogeneously by the whole material. He found that APE is independent of vibration frequency or the temperature of the material. But APE highly depends on the amplitude of ultrasound. After continuously testing of the effect of ultrasound energy level, Langenecker proposed that dislocation theory and stress wave propagation were not enough to describe the interactions between ultrasonic waves and the dislocations in metal crystalline grains. He, instead, proposed an energies superposition hypothesis. The energy required for plastic deformation is the summation of thermal energy of a dislocation, the energy produced through internal friction of oscillation dislocations, and the energy resulting from acoustic strain. Langenecker's work has a large impact on the following decades of research work on APE. He had some suggestions in the future research direction based on quantum mechanics and phonon-phonon and phonon-electron interaction. But his work stopped there. The idea of phonon concept and quantum mechanics may be necessary to understand acoustic softening. At his time, semi-conductor industry and photo-voltaic industry are not well developed. In fact, energy band gap theory used for building photo-voltaic solar cells is an energy method based on quantum mechanics. We will talk about it in the later section.

Series of APE tests were also reported by Kirchner et al \citep{kirchner_plastic_1985}. An hourglass shaped aluminum alloy was subjected to a cyclic loading of 20 KHz. A quasi-static compression is applied at strain rate of 0.0001/second. And a vibratory loading at 0.5, 1.0, 10., 50., 20 KHz are simultaneously applied to the metal samples. Their experimental results showed no frequency influence on the APE. No microscopic structural changes such as rearrangement of dislocation structures are observed.  The measured mean stress dropped to a lower value after imposing vibratory loading. The stress will return back to its original static stress-strain curve after cessation of vibrations.
  And they realized that it is very difficulty to measure stress inside the specimen because of inhomogeneous stress distribution.

Oscillatory unidirectional stresses experiments on crystals of KCl, NaCl, and NaBr by Ohgaku and Takeuchi \citep{ohgaku_blaha_1987} confirmed that the effect of temperature on the APE was negligible. The higher amplitudes of imposed ultrasonic vibrations led to reduction in “static-loading stress”. They claimed that the theorem of stress superposition cannot explain acoustic softening effect.  

Mao et al.\citep{mao_investigating_2020} studied acoustic softening in aluminum and its alloys. They found  that the reduction in flow stress is linearly proportional to the applied ultrasound intensity for all three types of aluminum alloys.

Contrast to acoustic softening, residual acoustic hardening may happen if ultrasound of sufficiently high-stress amplitude is applied on the crystalline materials in a long period and turned off \citep{fu_investigation_2022}. This residual hardening will show up as higher stress in stress-strain curve after ultrasound is stopped. It will finally get saturated even with very high amplitude ultrasound \citep{tyapunina_characteristics_1982, zhou_influence_2018}.  It is assumed that dislocations reach the maximum number of dislocations in the crystals that can be created by ultrasound.

Numerical simulation also becomes a powerful tool for investigate ultrasonic welding processing \citep{yao_modeling_2023, zhao_molecular_2022, kang_crystal_2022}. Finite element analysis has been widely used \citep{daud_modelling_2007, shao_modelling_2023}. 
In ultrasonic bonding processing, a large elasto-plastic deformation happens. This observation makes people to think the current solid mechanics theory which is mostly built on small deformation at Lagrange frame cannot be used. The fundamental theory from Eulerian frame may be more suitable. Therefore, conservation of mass, momentum, and energy needed be solved with finite elastic-plastic constitutive model or elastic-plastic kinetic evolution equations \citep{yu_numerical_2010,cai_acousto-plastic_2006}. Fundamental simulation of finite elastoplastic deformation was attempted by many researchers \citep{rubin_eulerian_2019}. The numerical simulation can clearly show nonlinear wave stress profile, density evolution, and velocity distribution during the ultrasonic welding processing. But the limitation is obvious. Numerical errors by discretization of governing equation and boundary condition could smash results. Artificial damping is added into numerical models to stabilize the computation. The influence of huge amount of defects existed in the materials can hardly be modeled by numerical methods. The elastic-plastic energy flow may not be correctly modeled by the current elastic-plastic constitutive models or elastic-plastic kinetic evolution equations. Mathematical modeling of acoustic softening is even more challenging. Based on existed crystal plasticity theory, many mathematical models were built to understand acoustic softening effect. For example, by adding a softening factor into plastic slip flow rule, Siddiq and Sayed \citep{siddiq_ultrasonic-assisted_2012, siddiq_acoustic_2011} built a model to relate yielding stress softening to ultrasonic intensity. However, most mathematical structures based on dislocation theory are fairly complex \citep{zhai_dislocation_2023}. Usually many differential equations and more than 20 parameters are needed to capture acoustic softening \citep{yao_acoustic_2012}. Nearly all mathematical models are considering to add an additional softening parameter to lower down the yielding stress.  In this paper, a new approach is presented to understand acoustic softening with the yielding stress ignored. Plasticity of materials will be determined by Young's modulus, defect density and defect velocity. The stress deviates from the linear relationship by defect density changes. Defect density is assumed to strongly depend on the number of phonons which is proportional to the amplitude of ultrasonic wave.  A concise crystal plastic equation is used to explain acoustic softening effect in section III. In section II, the wave-particle duality of light and sound is discussed. Particularly, particle treatment of ultrasound will be illustrated.

\section{Wave-Particle Duality of Light and Sound}
\label{s:photoelectric} \noindent
In the early 20th century, there was a big challenge in physics called ultraviolet catastrophe. That is, black body will emit infinity energy if wavelength of light decreases to the ultraviolet range. The challenge was solved by Max Planck. He assumed that the energy of light is not continuous and should be discretized. His idea was later adopted by Albert Einstein to understand photoelectric effect. At that time everyone thought that light is electromagnetic wave. Its motion should obey Maxwell's equation.  Albert Einstein, however, explained photoelectric effect by treating light as particles instead of wave. He stated that light should be viewed as wave in some cases and particles in the some other cases. When light is irradiating on a metal surface, electrons will absorb energy from light particles or photons. If photon energy is larger than the work function of the metal, electrons will be ejected from the metal. The number of electrons ejected is proportional to the number of photons. The photon energy is proportional to its frequency only, e.g., $E=h \nu$, where $E$ is the energy of photons, $h$ is Planck's constant, $\nu$ is frequency. Therefore, the maximum kinetic energy of the photoelectrons is independent of the light intensity. The kinetic energy of emitted electrons is determined by light frequency. On the other hand, the number of electrons ejected is proportional to the intensity of the incident light. Amplitude of electric current is proportional to available free electrons ejected by photons.  So if light is treated as wave, its energy is related to wave amplitude. If it is treated as photon, energy of photon is proportional to its frequency. Therefore, photon flux can be defined as the amount of photons per time per area. For fixed wave length,  photon flux $\Phi$ is related to light wave power density $P$ by
\begin{equation} 
\Phi \: = P \frac{\lambda}{h c} \,  
\label{eq:localized_3} 
\end{equation} 
where $c$ is the speed of light and $\lambda$ is the wave length of light.

\begin{figure}
\includegraphics[scale = 0.5]{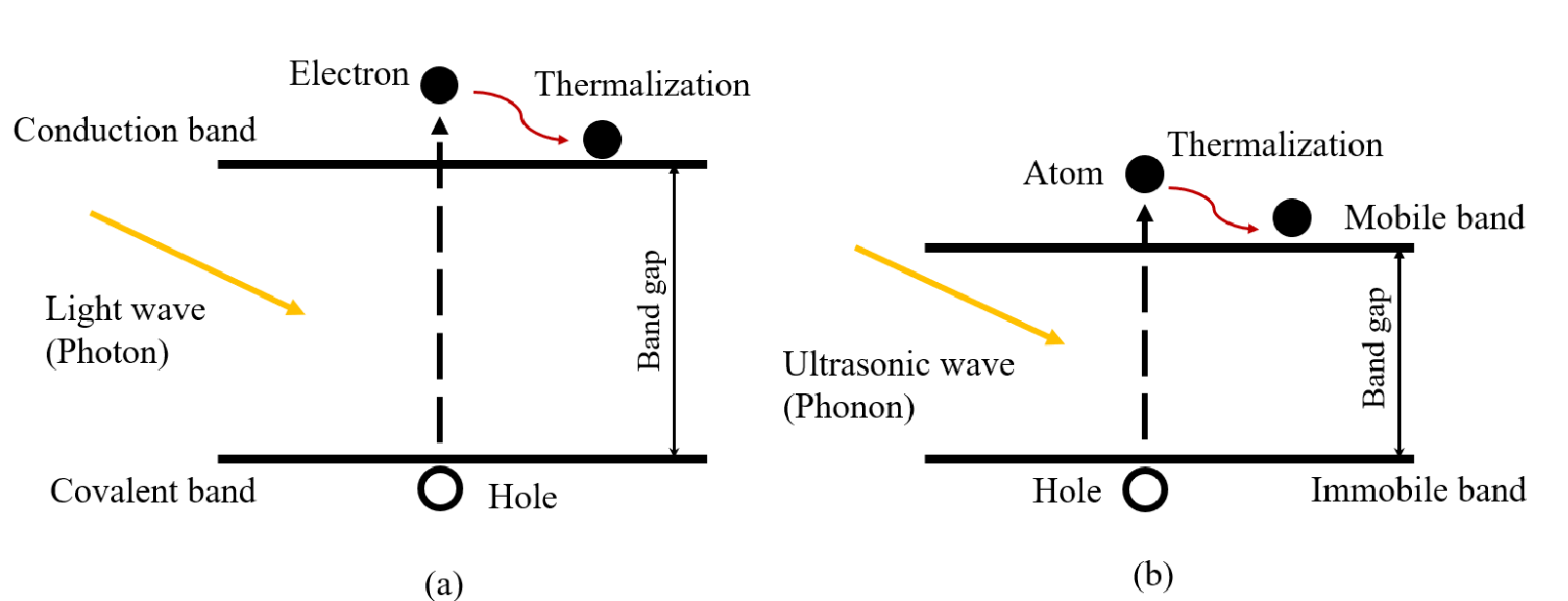}
\caption{
Schematic sketch of energy band gap theory. (a) Electrons in covalent band jump to conduction band after absorbing photon energy. (b) Defect atoms in immobile band jump to mobile band after absorbing phonon energy.
}
\label{f:band-gap}
\end{figure}

	The particle treatment of light also plays crucial role in photovoltaic industry. Similar to photoelectric effect, photovoltaic effect, firstly discovered by Edmond Becquerel in 1839, now has been utilized to produce solar cells. When high enough frequency light is shining on solar cells, photon energy with different frequency will be absorbed by different band gap semi-conductor materials. Photon energy will be absorbed by covalent electrons. After covalent electrons
gain enough energy from photon, for example, larger than band gap energy, they then have enough
energy to move away from the constraint of nuclei. They become free and move around inside
materials. In an energy concept, electrons will jump from covalent band to conduction band. A schematic sketch of band gap concept for electrons is shown in Fig.(\ref{f:band-gap})(a). 
When electrons jump to conduction band, holes will be created in covalent band. Incident light causes electron and hole pairs to be generated in the semiconductor. Electrons are free to move in conduction band and holes are free to move in covalent band. Both the conduction-band electrons and covalent band holes play important parts in the electrical behavior of photovoltaic cells. Electrons and holes freed from their positions in the crystal by light photon are called as light-generated excessive electron-hole pairs. If there is no internal built electric potential,
these light-generated excessive electrons and holes will recombine immediately when light is turned off, which means the energy of electrons and holes cannot be utilized. In solar cells, however, in order to use the electron and hole energy, a
build-in electric potential, e.g., PN junction (P type crystalline silicon and N type crystalline silicon junction) is needed. PN junction will separate electrons and holes. Electrons and holes will move in the opposite direction.
When electrons and holes move across an external circuit, their energy can be collected.  That is how solar cell creates power.

\begin{figure}
\includegraphics[scale = 0.6]{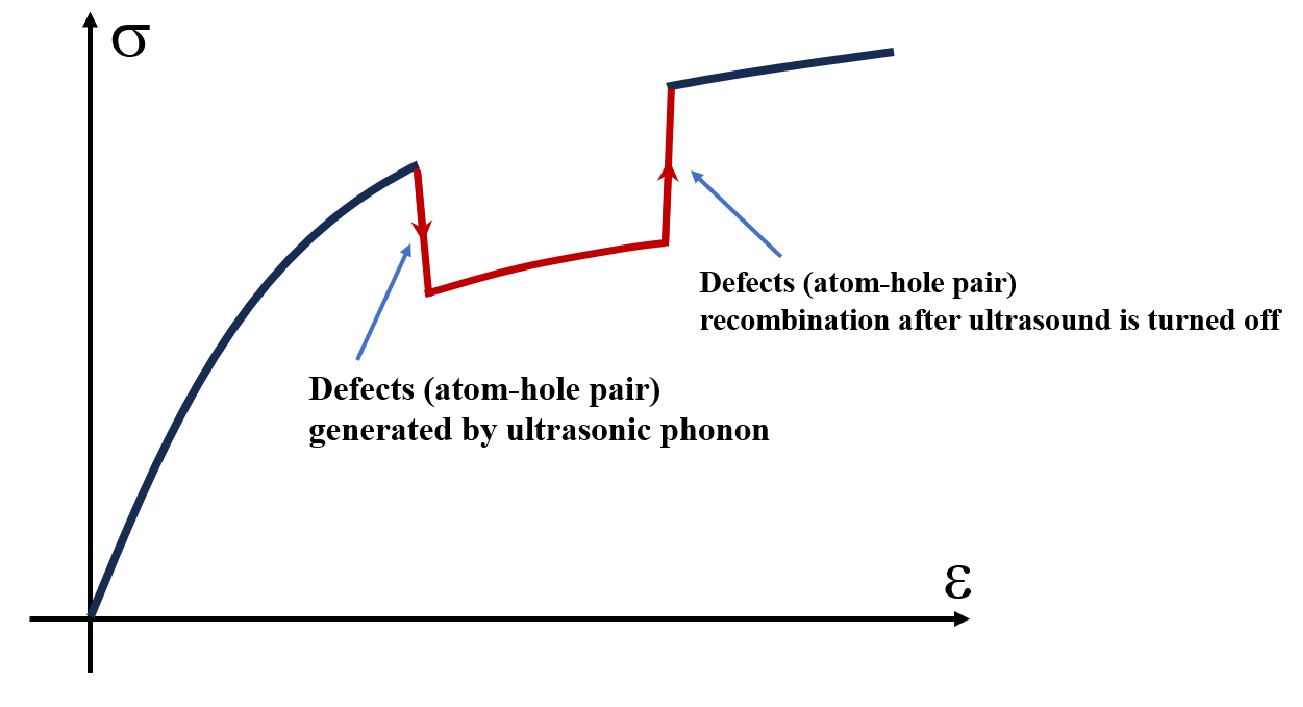}
\caption{
Schematic sketch of stress-strain curve with ultrasound turned on and turned off.
}
\label{f:SE-U}
\end{figure}

Similarly, there is an analogy between light wave and acoustic wave. They share the same wave characteristic properties such as wave reflection,  wave refraction and total internal reflection. When studying nano transport theory, Gang Chen \citep{chen_nanoscale_2005} treated electrons, photons, and phonons in a unified way.  Just like light wave, ultrasonic wave should have both wave and particle properties. Ultrasonic wave is a stress wave with frequency higher than 20000 Hz. It has all wave properties such as wave speed, wave length, amplitude and frequency. But its particle side is usually ignored by many researchers.  By analogy to photon in light wave, the particle treatment of acoustic wave is called phonon \citep{mason_phonon_1960}. Phonon is the smallest energy carrier of ultrasonic wave.
Just like photons can interact with electrons, phonons will interact with defects such as dislocations when ultrasonic wave is irradiated on materials. The interaction between ultrasonic wave and defects is very
complex. It will be related to inelastic scattering. However, if ultrasonic wave is treated as phonons, all
complexity can be solved. The power density or amplitude of ultrasonic wave is related to the number of
phonons. The phonon energy is given by $h\nu$, where $h$ is Planck's constant and $\nu$ is the frequency of ultrasonic wave. The number of mobile defects such as mobile dislocations generated by phonons is
proportional to the number of phonons. Since the mesoscale defects inside materials are generally due to missing or redundancy of atoms, these defects such as mobile dislocations here are treated as atom-hole pairs similar to electron-hole pairs generated by light wave. A similar energy band gap treatment of phonon induced atom-hole pair is sketched in Fig.(\ref{f:band-gap})b. After obtaining phonon energy, the weakly bonded atoms at defect area such as dislocation area will jump from immobile band to mobile band and leave holes behind in the immobile band. So atoms in mobile band and holes in immobile band can move around. Their movements will cause plastic deformation. In acoustic softening experiments, people found that the extent of stress softening of materials is proportional to the amplitude of the ultrasonic wave applied but independent of wave frequency. If ultrasound is treated as
particles, high amplitude ultrasonic wave means more phonons. It will create more mobile atom-hole pairs. The activation energy to to free atoms from their pinned equilibrium positions is the band gap energy which is the minimum energy to create mobile atom-hole pairs. Different materials have different band gap energy. Phonon energy need be larger than band gap energy to active dislocations and create mobile atom-hole pairs. Since phonon energy only depend on its frequency, high frequency phonon is more like to create atom-hole pairs.  The number of mobile atom-hole pairs is independent of frequency of ultrasonic wave. The kinetic energy of atom-hole pairs is decided by band gap energy. The excessive energy (the difference of phonon energy and band gap energy) absorbed by pinned dislocations will go through thermalization process and change to heat in a very short time. Therefore, defect density and defect velocity are independent of ultrasonic frequency. Defect density in materials will depend on the number of phonons or amplitude of ultrasonic wave. With more mobile atom-hole pairs generated by ultrasound, materials become easy to deform. If ultrasound is turned off, mobile atoms and holes will recombine since there is no built-in potential to separate atoms and holes like PN junction in semi-conductor materials. Once atoms and holes are recombined, less mobile atom-hole pairs will remain. Stress will return back to the original position. A sketched stress-strain curve with ultrasound is given by Fig.(\ref{f:SE-U}).

\begin{figure}
\includegraphics[scale = 0.5]{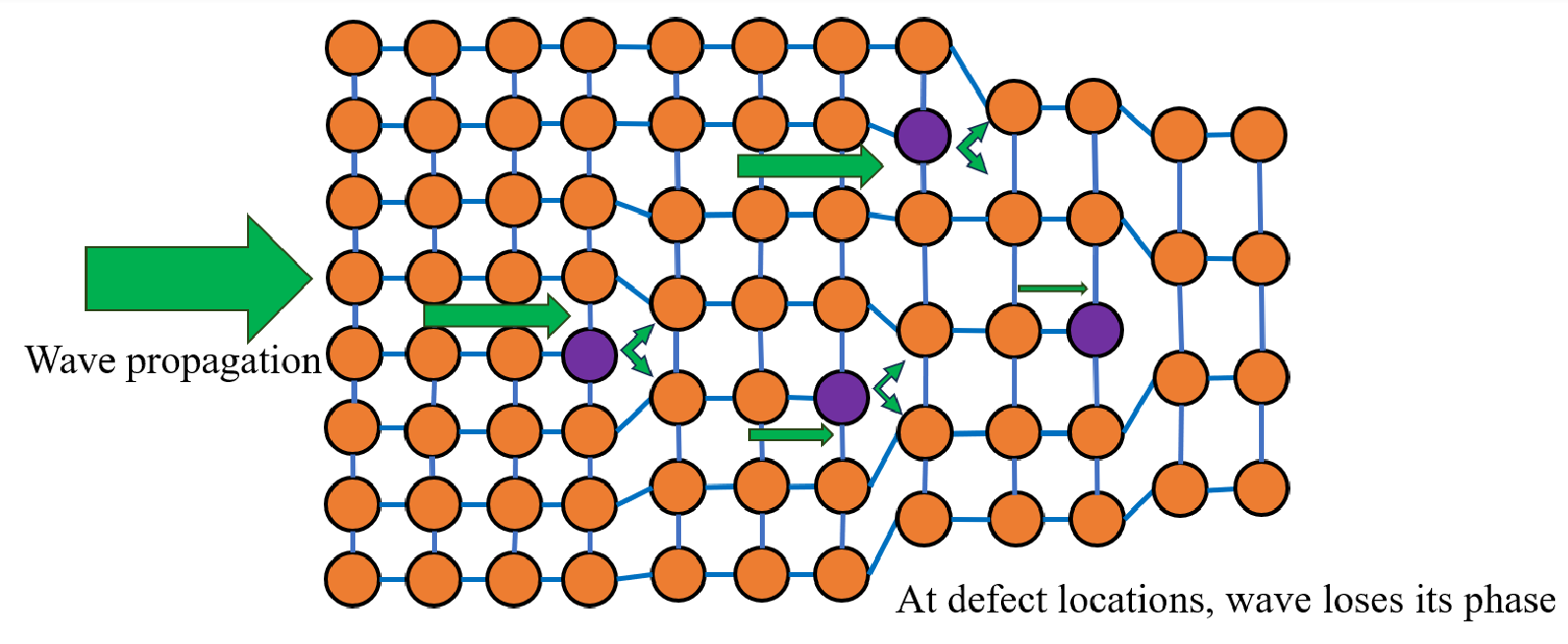}
\caption{
Schematic sketch of ultrasonic wave interaction with defects (purple color atoms). Ultrasonic wave loses its energy to defects and also loses its phase during its propagation.
}
\label{f:wave-particle-d}
\end{figure}

Both light and sound have wave and particle characterization.  In some cases they will act like wave. In some other cases they will act like particles. But how do we know it should be treat as wave or particles? Albert Einstein didn't give us an answer. One idea is that when wave packet size is much less than domain size, it be treated as particles. Gang Chen \citep{chen_nanoscale_2005} used coherent phase to explain wave-particle duality. He proposed that if an incident wave loses its phase during its propagation, it should be treated as particles. We agree with his idea. A propagating ultrasonic wave may easily lose its phase when it meets with defects. Its energy will be absorbed by pinned atoms and create mobile atom and hole pairs. These mobile atoms will move around and collide with lattices and lose their energy, which leads to thermalization. As a result, wave energy is lost at the defect areas. This can be sketched in Fig.(\ref{f:wave-particle-d}). So defects will act as an impedance or a barrier to acoustic wave propagation. If defect density is large enough, wave speed can be taken as zero for acoustic wave \citep{yang_revisit_2020}. This is because wave energy is all absorbed by defects and the phase of acoustic wave is entirely lost. Therefore, acoustic wave need be treated as phonon when there exists plenty of defects. This happens to electrons too. It is called quantum tunneling where kinetic energy of electrons is less than the potential barrier they are trying to pass through.   

\section{A Mathematical Model of Acoustic Softening}
\label{s:ultrasonic} \noindent

As derived for both amorphous and crystalline materials, the one-dimensional defect based elastic-plastic model can be written as \citep{yang_revisit_2020}
\begin{equation} 
\frac{d \sigma}{d \epsilon} \: = \: E \, - \, q \sigma^n \epsilon^m \,  
\label{eq:localized} 
\end{equation}
where  $\sigma$ is stress, $\epsilon$ is strain, $E$ is Young's modulus, $ q$ is a constant. m an n depend on defect density and defect velocity evolution respectively. 
In this stress-strain relationship, no yielding stress is needed. Defect density will gradually increase with increased external stress. The second term of Eq.(\ref{eq:localized}) will increase with stress and strain. The slope of stress and strain curve will gradually decrease and a maximum stress can be reached if slope becomes zero.

 In the derivation of Eq.(\ref{eq:localized}),  two formula are adopted:   defect density is related to strain by $\rho = a\epsilon^m$ and defect velocity is related to stress by $v = b \sigma^n$. With these two equations, Eq.(\ref{eq:localized}) can be rewritten as
\begin{equation} 
\frac{d \sigma}{d \epsilon} \: = \: E \, - \, C \rho v \,,  
\label{eq:localized_1} 
\end{equation}
where $C$ is a constant, $\rho$ and $v$ are defect density and defect velocity. Young's modulus, $E$, depends on time, temperature, hydrostatic pressure and immobile defect density \citep{yang_viscoelasticity_2021}. As long as velocity of defects inside materials is zero, the second term of Eq.(\ref{eq:localized_1}) is zero. Eq.(\ref{eq:localized_1}) will return back to Hooke's law. When ultrasonic wave is added, phonon will create excessive atom-hole pair. These atoms and holes will move around. Defect density will increase by additional term, e.g., $\delta \rho$. Therefore, stress and strain relationship without ultrasonic vibration, e.g., Eq. (\ref{eq:localized_1}), need be changed to

\begin{equation} 
\frac{d \sigma}{d \epsilon} \: = \: E \, - \, C (\rho + \delta \rho) v.  
\label{eq:localized_2} 
\end{equation}
Eq.(\ref{eq:localized_2}) is stress and strain relationship with ultrasonic vibration. An excessive defect density $\delta \rho$ is added to the equation.
This excessive defect density strongly relates to the amount of phonon, which is proportional to intensity of ultrasonic wave. Similar to photon flux,  phonon flux is defined as the amount of phonons per time per area. For fixed wave length,  phonon flux $\Phi$ is related to power density $P$ by
\begin{equation} 
\Phi \: = P \frac{\lambda}{h C_s} \,  
\label{eq:localized_3} 
\end{equation} 
where $h$ is Plank's constant, $C_s$ is speed of ultrasound. Unit of power density is given as watts per area.

\begin{figure}
\includegraphics[scale = 0.55]{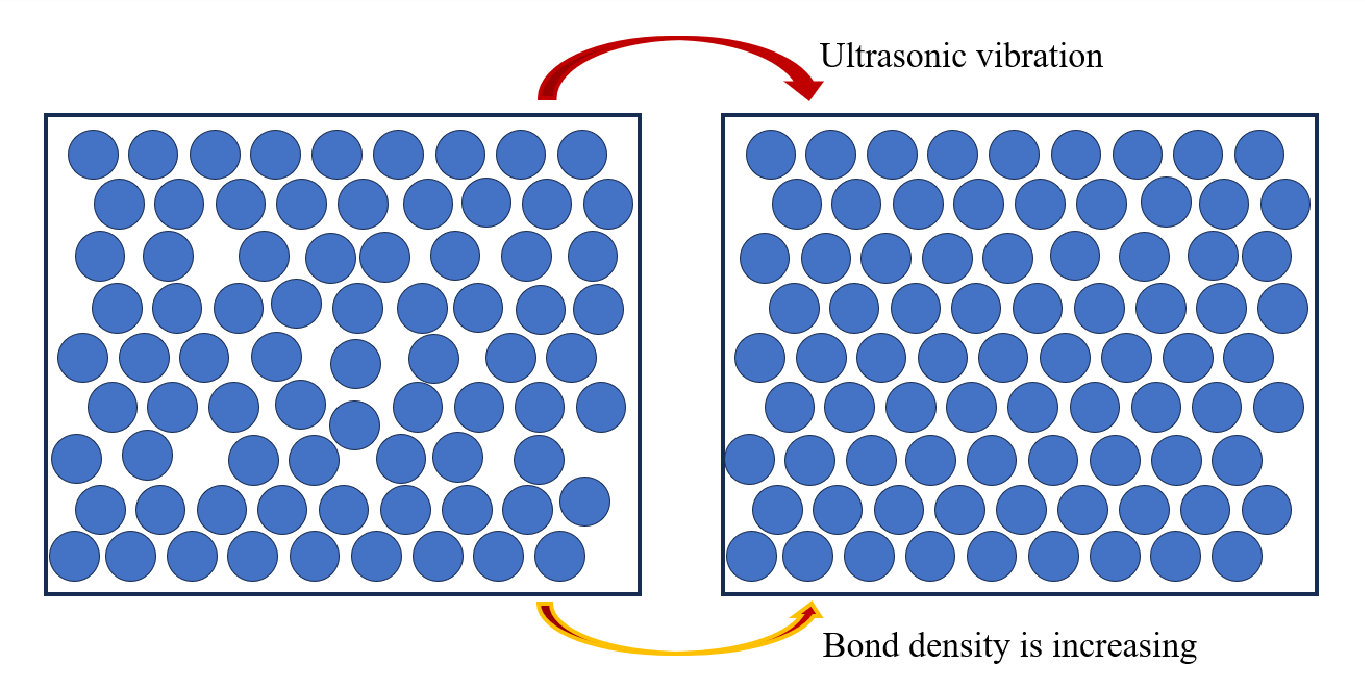}
\caption{
Atoms are more closely packed after long time ultrasonic vibration. Young's modulus will increase due to long time ultrasonic vibration. 
}

\label{f:density}
\end{figure}

Stress and strain relationship with ultrasonic vibration can also be written as the following form
\begin{equation} 
\frac{d \sigma}{d \epsilon} \: = \: E \, - \, q \sigma^n (\epsilon^m + \epsilon_U),  
\label{eq:localized_4} 
\end{equation}
where excessive strain, $\epsilon_U$, is due to excessive defect density $\delta \rho$ created by ultrasonic wave.
With higher intensity of ultrasonic wave, $\delta \rho$ is larger. Strain is also larger.  The slope of stress and strain curve will become smaller. Stress will become lower.
After ultrasound is turned off, atoms and holes will recombine immediately.  Therefore, excessive defect density, e.g., $\delta \rho$, will become zero. Stress-strain curve will return back to the original curve without ultrasound imposed. However, this is just an idealized case. Defect density and locations cannot be exactly the same as those before ultrasound is exerted on the materials. Depending on ultrasonic vibration time, the final immobile defect density left inside materials can be different compared to initial defect density. When vibration time is short, there is no enough time for atom-hole pairs to combine. There are more defects left in the materials. The atoms are less closely packed. Recall that Young's modulus is determined by number of bonding of atoms in a unit volume. So Young's modulus will decrease. This leads to residual acoustic softening effect. When vibration time is long enough, mobile atom-hole pairs have enough time to combine. Atoms in the materials are more closely packed. Bonding density of unit volume will increase. Young's modulus will got enhanced, which leads to residual acoustic hardening effect. Young's modulus variation by ultrasonic vibration was observed by several researchers \citep{zhen-yu_effect_2023}. This can be shown in our mathematical model in the next section. A schematic sketch of atom redistribution due to ultrasonic wave is shown in Fig.(\ref{f:density}).

\begin{figure}
\includegraphics[scale = 0.6]{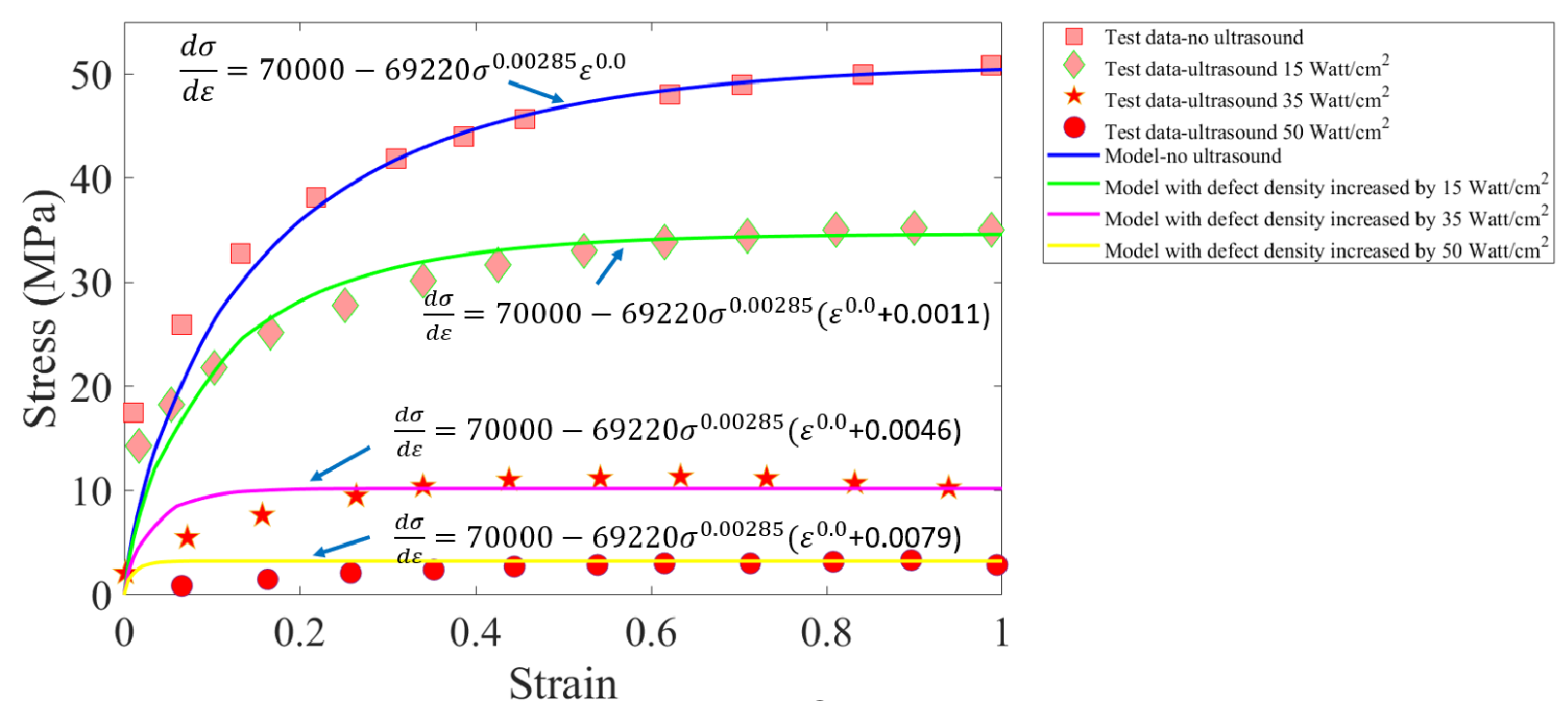}
\caption{
Experimental data by Langenecker \citep{langenecker_effects_1966} are compared to model prediction. Experimental data are represented by different shape and color markers. Model results are shown by using different color solid lines. The equations and parameters are also shown to predict stress and strain curves under no ultrasonic vibration (blue line), 15 $watts/cm^2$ (green line), 35 $watts/cm^2$ (magenta line), 50 $watts/cm^2$ (yellow line).  
}

\label{f:validation1}
\end{figure}

To validate the current mathematical model, model results are compared to two sets of experimental results. Firstly, the model is used to compare with stress and elongation curve of aluminum single crystals. The experiments were performed by Langenecker \citep{langenecker_effects_1966}. Ultrasonic wave with 20 k cycle/second was used in the experiments. He firstly got a stress-strain curve without ultrasonic vibration. Then he added ultrasonic wave to the tests. By increasing power density of ultrasonic wave from 15 $watts/cm^2$ to 50 $watts/cm^2$, he found the stress that leads to the same amount strain will decrease. Ultrasonic vibration makes materials easy to deform. His experimental data are replotted in Fig.(\ref{f:validation1}) to compare with our model prediction. In the model equation, Eq. (\ref{eq:localized_4}), Young's modulus of high-purity aluminum single crystals is set to be 70 GPa.  $q$ is choose to be 69.22 GPa. $m = 0.00285$ and $n =0$. Excessive strain, $\epsilon_U$, is 0 under no ultrasonic vibration.  Excessive strain, $\epsilon_U$ is selected to be 0.0011, 0.0046, 0.0079 for ultrasonic power density level 15 $watts/cm^2$, 35 $watts/cm^2$ and 50 $watts/cm^2$ respectively. In small strain, e.g., $\epsilon <0.2$, model prediction is a little deviation from the experimental data. As plastic strain increases, model prediction becomes better. The overall stress and strain shapes under different power density ultrasonic vibration are well predicted by our mathematical model. Since Eq.(\ref{eq:localized_4}) is a nonlinear equation, the relationship of stress and strain  is calculated by using nonlinear equation solver $ode23s$ in Matlab. In the calculation, stress has real part and imaginary part. Real part is used and imaginary part is discarded. 

\begin{figure}
\includegraphics[scale = 0.5]{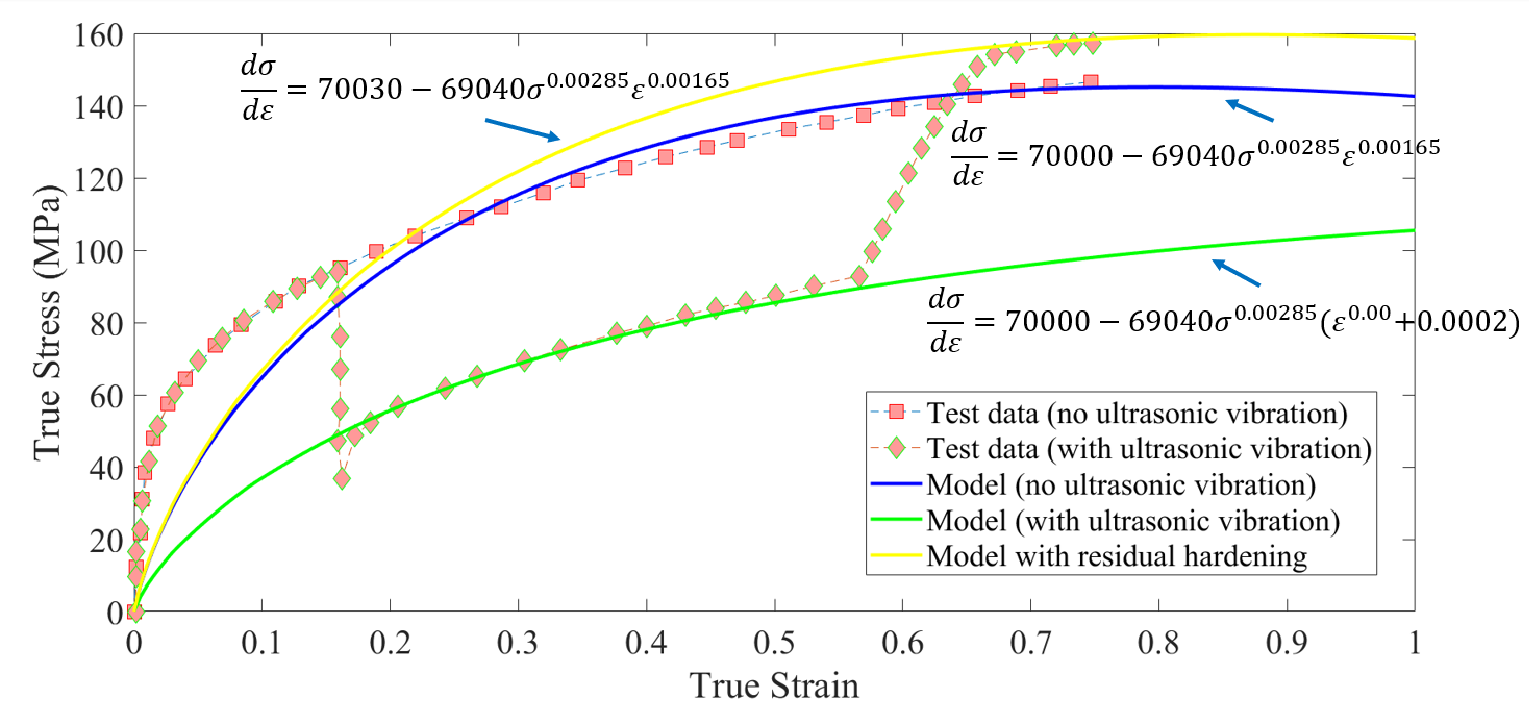}
\caption{
Model predictions are compared to experimental data conducted by Yao et al.\citep{yao_acoustic_2012}. Experimental data are represented by different shape and color markers. Model results are shown by using different color solid lines. Model predicted results are shown as blue line (no ultrasonic vibration), green line (with ultrasonic vibration) and yellow line (residual acoustic hardening).
}
\label{f:validation2}
\end{figure}

To further validate our model equation, model results are compared to another set of experimental data given by Yao \citep{yao_acoustic_2012}. Vibration frequency 9.6 $Hz$ was selected for the commercially pure aluminum (Al 1100, McMaster-Carr) ultrasonic vibration tests.  In the experimental tests, they observed that flow stress is significantly reduced once ultrasonic vibration is turned on. They also saw residual hardening of materials after ultrasonic vibration is turned off. In their tests, residual hardening can be seen when vibration time is 8 seconds and no residual hardening is observed when vibration time is 2 seconds. Their tests also showed that effect of strain rate on residual hardening is smaller than the effect of ultrasonic vibration time.  As we discussed before, after long time ultrasonic vibration, there is more time for atoms and holes to recombine. Young's modulus will increase by a small amount. In our equation, Eq.(\ref{eq:localized_4}), if let Young's modulus increase from initial value, $70000 MPa$ to a new value, $70030 MPa$, and keep all other parameters unchanged, flow stress will increase due to this small change of Young's modulus. Stress and strain curve with residual hardening obtained by the mathematical model is plotted against experimental data as shown in Fig. (\ref{f:validation2}). When strain is less than $0.1$, model results deviate from the experiments. When strain increases, especially, in large plastic zone, model prediction gets better. For small strain region, model prediction can be improved if many low defect energy terms are added to the right hand of Eq.(\ref{eq:localized_4}).

\section{Conclusion} 
\label{s:conclusion}
In the past, ultrasonic wave propagation theory and dislocation theory have been continually used to understand acoustic softening. After many years' development, more and more complex physical concepts and mathematical structures are added. This made the science behind of acoustic softening even elusive. In this paper, however, some attempts to simplify its physical concepts and mathematical structures are presented. The ultrasonic wave is treated as particles, e.g., phonons. We walk away from traditional dislocation theory by introducing energy band gap theory used in semi-conductor physics.

Energy of phonon is assumed to be proportional to its frequency. Number of defects created by ultrasonic wave is proportional to phonon numbers or power density of ultrasonic wave. Residual softening or hardening after ultrasonic vibration is turned off is related to the change of Young's modulus. 

Energy band gap theory is originally built to explain electrons jump from covalent band to conduction band after light irradiation. This theory is extended to explain atoms jump from immobile band to mobile band after ultrasonic wave irradiation. Based on atom-hole pair generation and recombination,  ultrasonic wave induced excessive defect density is added to a crystal plasticity equation. Yielding stress is not required in this mathematical model. The new constitutive model is bench marked with two experimental results and shows good consistence.

\section*{Data Availability}
Data sharing not applicable - no new data generated.
\section*{Reference}
\bibliographystyle{elsarticle-num}
\bibliography{zhou_dft}

\begin{thebibliography}{10}
\expandafter\ifx\csname url\endcsname\relax
  \def\url#1{\texttt{#1}}\fi
\expandafter\ifx\csname urlprefix\endcsname\relax\def\urlprefix{URL }\fi
\expandafter\ifx\csname href\endcsname\relax
  \def\href#1#2{#2} \def\path#1{#1}\fi

\bibitem{cai_acousto-plastic_2006}
M.~Cai, Acousto-{Plastic} deformation of metals by nonlinear stress waves,
  Ph.D. thesis, The Ohio State University (2006).

\bibitem{siddiq_theoretical_2009}
A.~Siddiq, E.~Ghassemieh, Theoretical and {FE} {Analysis} of {Ultrasonic}
  {Welding} of {Aluminum} {Alloy} 3003, Journal of Manufacturing Science and
  Engineering 131~(041007).

\bibitem{hu_impact_2017}
J.~Hu, T.~Shimizu, M.~Yang, Impact effect of superimposed ultrasonic vibration
  on material characteristics in compression tests, Procedia Engineering 207
  (2017) 1063--1068.

\bibitem{gallego-juarez_power_2023}
J.~A. Gallego-Juarez, K.~F. Graff, M.~Lucas (Eds.), Power {Ultrasonics}:
  {Applications} of {High}-{Intensity} {Ultrasound}, 2nd Edition, Woodhead
  Publishing, 2023.

\bibitem{deng_study_2023}
T.~Deng, H.~Liu, A study of mechanical characteristics and microstructural
  evolution of copper-nickel alloy sheet undergoing ultrasonic vibration
  assisted uniaxial tension, Materials Science and Engineering: A 885 (2023)
  145608.

\bibitem{baker_dislocation_2004}
G.~S. Baker, S.~H. Carpenter, Dislocation {Mobility} and {Motion} under
  {Combined} {Stresses}, Journal of Applied Physics 38~(4) (2004) 1586--1591.

\bibitem{izumi_effects_1966}
O.~Izumi, K.~Oyama, Y.~Suzuki, Effects of {Superimposed} {Ultrasonic}
  {Vibration} on {Compressive} {Deformation} of {Metals}, Transactions of the
  Japan Institute of Metals 7~(3) (1966) 162--167.

\bibitem{sedaghat_ultrasonic_2019}
H.~Sedaghat, W.~Xu, L.~Zhang, Ultrasonic vibration-assisted metal forming:
  {Constitutive} modelling of acoustoplasticity and applications, Journal of
  Materials Processing Technology 265 (2019) 122--129.

\bibitem{wang_acoustic_2016}
C.~J. Wang, Y.~Liu, B.~Guo, D.~B. Shan, B.~Zhang, Acoustic softening and stress
  superposition in ultrasonic vibration assisted uniaxial tension of copper
  foil: {Experiments} and modeling, Materials \& Design 112 (2016) 246--253.

\bibitem{cheng_acoustic_2023}
R.~Cheng, S.~Rose, A.~Taub, Acoustic softening – investigation of the volume
  effect and introduction of amplitude strain parameter, Materials Science and
  Engineering: A 881 (2023) 145437.

\bibitem{nevill_effect_1957}
G.~E. Nevill, Effect of vibrations on the yield strength of a low carbon steel,
  Ph.D. thesis, Duke University (1957).

\bibitem{kirchner_plastic_1985}
H.~O.~K. Kirchner, W.~K. Kromp, F.~B. Prinz, P.~Trimmel, Plastic deformation
  under simultaneous cyclic and unidirectional loading at low and ultrasonic
  frequencies, Materials Science and Engineering 68~(2) (1985) 197--206.

\bibitem{blaha_plastizitatsuntersuchungen_1959}
F.~Blaha, B.~Langenecker, Plastizitätsuntersuchungen von metallkristallen in
  ultraschallfeld, Acta Metallurgica 7~(2) (1959) 93--100.

\bibitem{langenecker_effects_1966}
B.~Langenecker, Effects of {Ultrasound} on {Deformation} {Characteristics} of
  {Metals}, IEEE Transactions on Sonics and Ultrasonics 13~(1) (1966) 1--8.

\bibitem{ohgaku_blaha_1987}
T.~Ohgaku, N.~Takeuchi, The {Blaha} {Effect} of {Alkali} {Halide} {Crystals},
  physica status solidi (a) 102~(1) (1987) 293--299.

\bibitem{mao_investigating_2020}
Q.~Mao, N.~Coutris, H.~Rack, G.~Fadel, J.~Gibert, Investigating
  ultrasound-induced acoustic softening in aluminum and its alloys, Ultrasonics
  102 (2020) 106005.

\bibitem{fu_investigation_2022}
Z.~Fu, G.~Gao, Y.~Wang, D.~Wang, D.~Xiang, B.~Zhao, Investigation of acoustic
  softening and microstructure evolution characteristics of {Ti3Al}
  intermetallics undergoing ultrasonic vibration-assisted tension, Materials \&
  Design 222 (2022) 111015.

\bibitem{tyapunina_characteristics_1982}
N.~A. Tyapunina, V.~V. Blagoveshchenskii, G.~M. Zinenkova, Y.~A. Ivashkin,
  Characteristics of plastic deformation under the action of ultrasound, Soviet
  Physics Journal 25~(6) (1982) 569--578.

\bibitem{zhou_influence_2018}
H.~Zhou, H.~Cui, Q.~H. Qin, Influence of ultrasonic vibration on the plasticity
  of metals during compression process, Journal of Materials Processing
  Technology 251 (2018) 146--159.

\bibitem{yao_modeling_2023}
Q.~Yao, J.~Yang, P.~Dong, Z.~Zhao, Y.~He, Y.~Ma, F.~Zhang, W.~Li, Modeling of
  acoustic field dependent tensile property for metal materials, Extreme
  Mechanics Letters 60 (2023) 101980.

\bibitem{zhao_molecular_2022}
Y.~Zhao, J.~Zhai, Y.~Guan, F.~Chen, Y.~Liu, Y.~Li, J.~Lin, Molecular dynamics
  study of acoustic softening effect in ultrasonic vibration assisted tension
  of monocrystalline/polycrystalline coppers, Journal of Materials Processing
  Technology 307 (2022) 117666.

\bibitem{kang_crystal_2022}
J.~Kang, X.~Liu, S.~R. Niezgoda, Crystal plasticity modeling of ultrasonic
  softening effect considering anisotropy in the softening of slip systems,
  International Journal of Plasticity 156 (2022) 103343.

\bibitem{daud_modelling_2007}
Y.~Daud, M.~Lucas, Z.~Huang, Modelling the effects of superimposed ultrasonic
  vibrations on tension and compression tests of aluminium, Journal of
  Materials Processing Technology 186~(1) (2007) 179--190.

\bibitem{shao_modelling_2023}
G.~Shao, H.~Li, X.~Zhang, J.~Zou, Z.~Huang, M.~Zhan, Modelling of ultrasonic
  vibration-assisted forming considering the distribution of ultrasonic field
  with structure deformation, International Journal of Plasticity 170 (2023)
  103744.

\bibitem{yu_numerical_2010}
S.-T.~J. Yu, L.~Yang, R.~L. Lowe, S.~E. Bechtel, Numerical simulation of linear
  and nonlinear waves in hypoelastic solids by the {CESE} method, Wave Motion
  47~(3) (2010) 168--182.

\bibitem{rubin_eulerian_2019}
M.~B. Rubin, An {Eulerian} formulation of inelasticity: from metal plasticity
  to growth of biological tissues, Philosophical Transactions of the Royal
  Society A: Mathematical, Physical and Engineering Sciences 377~(2144) (2019)
  20180071.

\bibitem{siddiq_ultrasonic-assisted_2012}
A.~Siddiq, T.~El~Sayed, Ultrasonic-assisted manufacturing processes:
  {Variational} model and numerical simulations, Ultrasonics 52~(4) (2012)
  521--529.

\bibitem{siddiq_acoustic_2011}
A.~Siddiq, T.~El~Sayed, Acoustic softening in metals during ultrasonic assisted
  deformation via {CP}-{FEM}, Materials Letters 65~(2) (2011) 356--359.

\bibitem{zhai_dislocation_2023}
M.~Zhai, C.~Wu, L.~Shi, G.~Chen, Q.~Shi, Dislocation strain energy based
  modeling for ultrasonic effect on friction stir lap welding process of
  dissimilar {Mg}/{Al} alloys, Journal of Materials Research and Technology 22
  (2023) 252--268.

\bibitem{yao_acoustic_2012}
Z.~Yao, G.-Y. Kim, Z.~Wang, L.~Faidley, Q.~Zou, D.~Mei, Z.~Chen, Acoustic
  softening and residual hardening in aluminum: {Modeling} and experiments,
  International Journal of Plasticity 39 (2012) 75--87.

\bibitem{chen_nanoscale_2005}
G.~Chen, Nanoscale {Energy} {Transport} and {Conversion}: {A} {Parallel}
  {Treatment} of {Electrons}, {Molecules}, {Phonons}, and {Photons},
  illustrated edition Edition, Oxford University Press, Oxford ; New York,
  2005.

\bibitem{mason_phonon_1960}
W.~P. Mason, Phonon {Viscosity} and {Its} {Effect} on {Acoustic} {Wave}
  {Attenuation} and {Dislocation} {Motion}, Acoustical Society of America
  Journal 32 (1960) 458.

\bibitem{yang_revisit_2020}
L.~Yang, L.~Yang, Revisit initiation of localized plastic deformation: {Shear}
  band \& necking, Extreme Mechanics Letters 40 (2020) 100914.

\bibitem{yang_viscoelasticity_2021}
L.~Yang, L.~Yang, R.~L. Lowe, A viscoelasticity model for polymers: {Time},
  temperature, and hydrostatic pressure dependent {Young}'s modulus and
  {Poisson}'s ratio across transition temperatures and pressures, Mechanics of
  Materials 157 (2021) 103839.

\bibitem{zhen-yu_effect_2023}
Z.~Zhen-yu, Z.~Qiu-yang, L.~Yu, J.~Zhi-Guo, Y.~Zhi-peng, P.~Zhong-yu, Effect of
  ultrasonic vibration on the deformation through indentation test and
  molecular dynamics simulation, Mechanics of Materials 184 (2023) 104744.

\end{thebibliography}

%

\end{document}